\documentclass[preprint]{aastex61}
\begin{document}

\shortauthors{Breiding et al.}{}
\shorttitle{Fermi Flare Seed Photons from sheath illumination of the MT}

\title{Blazar Sheath Illumination of the Outer Molecular Torus: a Resolution of the Seed Photon Problem for the far-GeV Blazar Flares}

\author{Peter Breiding}
\affil{Department of Physics, University of Maryland Baltimore County, Baltimore, MD 21250, USA}

\author{Markos Georganopoulos}
\affil{Department of Physics, University of Maryland Baltimore County, Baltimore, MD 21250, USA}
\affil{NASA Goddard Space Flight Center, Code 660, Greenbelt, MD 20771, USA}
\correspondingauthor{Markos Georganopoulos} \email{georgano@umbc.edu}

\author{Eileen T. Meyer}
\affil{Department of Physics, University of Maryland Baltimore County, Baltimore, MD 21250, USA}

\begin{abstract}
Recent multiwavelength work led by the Boston University blazar group \cite[e.g.,][]{marscher10} strongly suggests that a fraction of the blazar flares  seen by the {\sl Fermi} Large Area Telescope (LAT) take place a few to several pc away from the central engine. However, at such distances from the central engine, there is no adequate external photon field to provide the seed photons required for producing the observed GeV emission under leptonic inverse Compton (IC) models. A possible solution is a spine-sheath geometry  for the emitting region (\citealt{macdonald15}, but see  \citealt{nalewajko14}). 
Here
we use the  current  view of the molecular torus \cite[e.g.,][]{elitzur12,netzer15} in which the torus extends a few pc beyond the dust sublimation radius with dust clouds distributed with a declining density for decreasing polar angle. We show
that for a spine-sheath blazar jet embedded in the torus,
the wide beaming pattern of the synchrotron radiation of the relatively slow sheath   will heat  molecular clouds whose subsequent IR radiation will be seen highly boosted in the spine comoving frame, and that under reasonable conditions this photon field  can dominate over the sheath photons directly entering the spine. If the sheath is sufficiently luminous it will sublimate the dust, and if the sheath synchrotron radiation extends to optical-UV energies (as  may happen during  flares), this will illuminate the sublimated dust clouds to produce emission lines that will vary in unison with the optical-UV continuum, as has been very recently reported for  blazar CTA 102 \citep{jorstad17}. 
\end{abstract}


\section{Introduction \label{section:introduction}}

Blazars, radio loud active galactic nuclei (AGN) with their relativistic jets pointed at small angles to the line of sight  \citep{blandford78},   exhibit powerful flares of  gamma-ray emission \citep[e.g.][]{abdo10}.  In the context of leptonic models, the gamma-ray emission of powerful blazars
is considered to be due to inverse Compton (IC) scattering  of low-energy seed photons
from the sub-pc size broad-line region  \cite[BLR;][]{sikora94} and/or the pc-scale molecular torus \citep[MT, ][]{blazejowski00}. 
These seed photons are in turn produced  when the semi-isotropically emitting  accretion disk  illuminates  the BLR and MT.

A related  seed photon production mechanism that motivated this work  is  the illumination of  BLR clouds by the beamed optical-UV radiation of the blazar itself \citep{ghisellini96}. 
Because the blazar emission is beamed within a small angle  ($ \sim 1/\Gamma \lesssim 5^\circ$), where $\Gamma\gtrsim 10$ is the bulk Lorentz factor of the flow), this model requires the existence
of  BLR clouds within this very small polar beaming angle.

The above  mechanisms posit  that the GeV emitting region is within  a pc or less from the central engine.   However,  multiwavelength monitoring \cite[e.g.][]{marscher10}, including VLBI imaging, reveals that at least a fraction of flares occur at the location of the VLBI core.
The inference that some gamma-ray flares originate near the location of the VLBI core results from the strong correlations seen in light curves from radio through gamma-ray wavelengths during these flares, and the fact that often  a superluminal component is seen ejected from the VLBI core near the peak time of the GeV  flare: in PKS 1510-089, one of the most well documented cases \citep{marscher10}, a radio and optical flare takes place at the same time with a  GeV flare and a new  component emerges from the VLBI core. The core  may be the site of a standing shock few to several pc downstream of the central engine. Alternatively, it can be the location where the jet at the frequency of observation becomes optically thick, again few to several pc downstream of the central engine. One can discriminate between these two possibilities,
as in the first case the location of the core does not change with the frequency of VLBI observation, while in the second it shifts closer to the central engine as the frequency of observation increases \citep[e.g.,][]{hada11}.

The far GeV flares  show that powerful blazar jets  produce gamma-ray emission beyond the canonical sub-pc scale BLR and pc-scale MT. At such large distances, however,  there is no substantial source of seed photons for  the IC process from the BLR or the MT, as both photon fields illuminate the blazar from behind and are therefore substantially  de-beamed   in the comoving frame of the emitting plasma \citep{dermer92, nalewajko14}.
An alternative source of seed photons arises if we consider that the jet is characterized by a fast spine in which the plasma flows with a Lorentz factor of the order of $\Gamma_{sp}\sim10-20$, surrounded  by a slow sheath through which the plasma flows with  a Lorentz factor of the order of $\Gamma_{sh}\sim$ few \citep{ghisellini05}. 
 In this case the energy density of the sheath photons will be seen to be Doppler boosted   in the spine  comoving frame by $\sim \Gamma_{sp}^2/(4 \Gamma_{sh}^2$) \citep{nalewajko14}.
\cite{macdonald15} applied the spine-sheath model to the blazar PKS 1510-089 and showed that with a judicious choice of model parameters it   can reproduce the observed variability.
 However,  \cite{nalewajko14} argued that the spine-sheath boosting is  inadequate for producing the required seed photon energy density in the frame of the spine without requiring a sheath spectral energy distribution  (SED) that overproduces the observed blazar SED. 
 
  Our current view of the MT suggests  another source of  seed photons.  Originally, the AGN unification scheme \citep[e.g.][]{antonucci93}   suggested that  the dichotomy between type 1 and type 2 AGN spectra can be explained as an orientation effect due to   an opaque equatorial MT. More recent observational and theoretical work \citep[e.g.][]{nenkova08,elitzur12,netzer15} support the idea that the MT is clumpy \citep{krolik88} and, furthermore, that the distribution of the dusty clumps is a gradually declining function of decreasing polar angle, allowing for some fraction of dusty clumps to lie at relatively small polar angles  \citep[e.g.,][]{garcia17,khim17}.  
  The inner radius $R_d$  of the MT is  set by  dust sublimation due to the  radiation of the accretion disk and has been measured by near-IR interferometry  \citep[e.g.,][]{kishimoto11} and near IR reverberation mapping \citep[e.g.,][]{koshida14} of nearby Seyferts to have a size that for bright AGN  (accretion disk luminosity $L_{optical-UV}\sim 10^{46}$ erg $s^{-1}$) would be $R_d\sim 1 $ pc \citep{nenkova08}.  
  
  The BLR  is contained within the dust sublimation radius $R_d$.  The outer radius of the MT is  found both with modeling the MT  SED \citep{fuller16} and through  Atacama Large Millimeter/sub-millimeter aray
 ({\sl ALMA}) observations \citep{garciaburillo16,gallimore16}  to be of the order of a few to several pc, several times larger than $R_d$,  with most of the MT power emitted in the IR from the clouds at $\sim R_d$ \citep{nenkova08}. 
 With this configuration it is plausible that the blazar radio core, identified with the blazar GeV flaring site  \citep[e.g.][]{marscher10} is located beyond $R_d$ but  within the outer bounds of the MT.

 Here we show that in the context of a spine-sheath configuration, a moderately relativistic sheath  located within the MT beyond $R_d$
  can illuminate a substantial solid angle of the clumpy dusty material that lies beyond it, out to the outer bounds of the MT. In  \S \ref{section:model} we show that the radiation of these illuminated clumps, seen  in the frame of the fast spine, can dominate over the sheath  radiation in the spine co-moving frame and act as the  required external photon field. In \S \ref{section:SED} we present a general model SEDs using a motivated set of model parameters,  and in \S\ref{section:discussion} we present our conclusions and discuss some of the implications of our model.

\section{The photon energy density in the jet spine}\label{section:model}

 \begin{figure*}[h]
 	\epsscale{1.1}
	\plotone{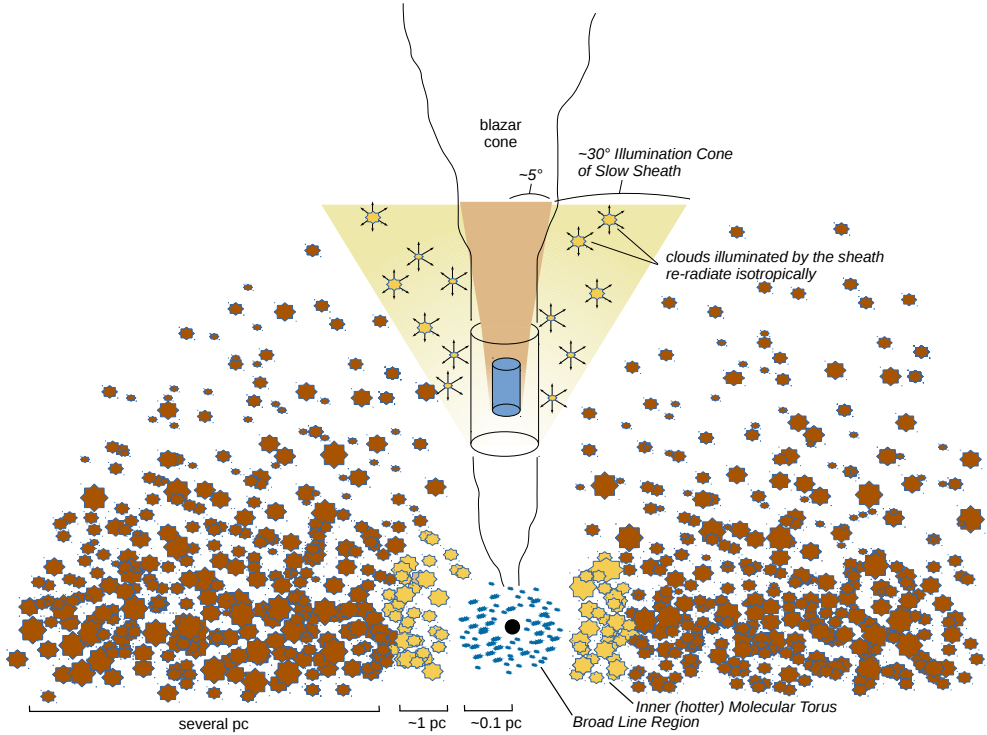}
	\caption{A schematic representation of the spine-sheath blazar embedded in the MT (not to scale). The MT is clumpy and the  clouds have
    a number density that  declines with decreasing polar angle. The blazar site is far beyond its traditional location within the BLR and/or the inner, hotter part of the MT: it is at a distance of a few pc but still within the MT. The wide beaming angle of its sheath synchrotron emission  illuminates and heats the MT clouds within the Sheath's synchrotron emission opening angle and  these  in turn radiate isotropically. This radiation can provide the dominant seed photon field  for IC scattering in the spine comoving frame. Note that the spine beaming angle is very small ($\lesssim 5^o$)  and it is not expected to be intercepted by any MT clouds.}
	\label{fig:schematic}
 \end{figure*}
 
Our picture (figure \ref{fig:schematic}) for the far GeV blazar emission posits a standing jet feature, possibly a recollimation shock, a few to several pc from the black hole. Plasma flows through the standing feature, which has a  spine  bulk Lorentz factor $\Gamma_{sp}\sim 10-20$ and a slower outer sheath with bulk Lorentz factor $\Gamma_{sh}\sim 2-4$.
To keep the study analytically tractable we approximately assume $1 \ll \Gamma_{sh}\ll \Gamma_{sp}$.  
Dusty molecular clouds within the wide (up to $\theta_{sh} \sim 1/\Gamma_{sh}\lesssim 30^\circ$) beaming angle of the synchrotron-emitting sheath  reprocess a fraction of this radiation, which is then relativistically amplified in the fast spine rest frame. The observed GeV emission is attributed to IC scattering of these photons by the relativistic electrons in the fast spine, provided this photon energy density dominates over that of the sheath photon field directly entering the spine.

We assume that the sheath produces an isotropic synchrotron luminosity $L_{sh}'$ in its comoving frame, peaking in the sub-mm to IR, as the synchrotron SEDs of powerful blazar usually do \cite[e.g.][]{meyer11}. In the galaxy frame, the solid-angle integrated
luminosity (the luminosity that a hypothetical detector covering all $4\pi$ of the sky of the source would measure) is
$L_{sh}=\Gamma_{sh} L'_{sh}$, valid assuming that the sheath is a stationary feature. Most of this radiation is beamed into a solid angle $\Omega_{sh} =\pi /\Gamma_{sh}^2$ (opening half-angle $\sim 1/\Gamma_{sh}$) and for simplicity we assume that within this angle the intensity of the radiation does not vary. An observer within this solid angle that assumes that the source is isotropic in her/his frame will infer a luminosity
 \begin{equation}
 L_{sh, obs}= L_{sh} 4\pi/\Omega_{sh}=4 \Gamma_{sh}^2  L_{sh}= 4 \Gamma_{sh}^3 L_{sh}'.
 \label{eq:lums}
 \end{equation} 
The  sheath synchrotron radiation illuminates MT dust clouds, which for simplicity we assume are isotropically and homogenously distributed  within the solid angle $\Omega_{sh}$, starting from the sheath radius $R_{sh}$  and extending to some distance $R_{out}$ with a sheath covering factor $C$
(where $C$ is the fraction of obscured solid angle within the sheath beaming cone).
We treat the dust clouds as ideal spherical black bodies which act as perfect absorbers to the sheath emission.  The dust clouds then absorb and are heated by the sheath radiation. Sublimation of the dust occurs for those clouds that are at distances from the sheath less than their sublimation radius 
\begin{equation}
R_{sub}=\left(\frac{L_{sh, obs}}{16 \pi \sigma T_{sub}^{4}}\right)^{1/2},
\label{eq:rsub}
\end{equation}
where $T_{sub}$ is the dust sublimation temperature, and $\sigma=5.67 \times 10^{-5}$ erg cm$^{-2}$ deg$^{-4}$ s$^{-1}$ is the Stefan-Boltzmann constant. The minimum $L_{sh,obs}$ that can cause sublimation is found by the requirement $R_{sub}=R_{sh}$ and is 
\begin{equation}
L_\star=16 \pi \sigma T_{sub}^4 R_{sh}^2=2.85\times 10^{45} \, T_{sub, 3}^4 R^2_{sh,18} \; {\rm erg} \, {\rm    s}^{-1},
\label{eq:lstar}
\end{equation}
where $T_{sub,3}$ is the dust sublimation temperature in units of $1000$ K, and $R_{sh,18}$ is the sheath radius in units of $10^{18}$ cm.
For $L_{sh,obs}<L_\star$ the sheath radiation heats but does not sublimate the dust, while for  $L_{sh,obs}>L_\star$ the dust is sublimated
up to a distance $R_{sub}$ given by equation (\ref{eq:rsub}). As we require that the observed blazar SED is dominated by the spine and not by the sheath, and as the typical synchrotron SED for powerful blazars has an observed power of $\sim 10^{46} \; {\rm erg} \, {\rm    s}^{-1}$ \cite[e.g.][]{meyer11}, we assume here that $L_{sh,obs}<L_\star$ (i.e. $\eta\equiv L_{sh, obs}/L_*<1$) and no sublimation takes place. 

\subsection{Photon energy density in the spine due to sheath radiation  reprocessed by the clouds}

We proceed now to derive the intensity of the radiation  received back from the dust clouds within $\Omega_{sh}$, and from this the corresponding photon energy density in the co-moving frame of the fast spine. 
 For this we assume that all of the sheath power  absorbed by the clouds found at  $R_{sh} < R < R_{out}$  and within $\Omega_{sh}$ is isotropically re-radiated as  thermal black body radiation for the dust clouds .

At a distance $r$ from the center of the spine-sheath system, the sheath power absorbed by a  shell of differential width $dr$ is
\begin{equation}
dP_{abs}=\frac{L_{sh} C}{R_{out}-R_{sh}}dr=  \frac{ L_{sh, obs} \Omega_{sh} C}{4\pi(R_{out}-R_{sh})}dr. 
\end{equation}
This power is then re-radiated by each shell with an emissivity given by: 
\begin{equation}
\epsilon=\frac{dP_{abs}}{dV}=\frac{dP_{abs}}{\Omega_{sh}r^{2}dr}=\frac{L_{sh, obs}C}{(R_{out}-R_{sh})4\pi r^{2}}.
\end{equation}
Assuming isotropic emission for each shell, the emission coefficient, $J$, can be written as $J=\epsilon/4\pi$. The intensity contribution from each infinitesimal shell is then given by $dI(r)=Jdr$, and the total intensity of the radiation field received back from the dust clouds is 
\begin{equation}
I=\int_{R_{sh}}^{R_{out}}J dr=\frac{L_{sh, obs}C}{16\pi^{2}R_{out}R_{sh}}=\frac{L_{sh, obs}C}{16\pi^{2}a R_{sh}^2},
\end{equation}
where $a\equiv R_{out}/R_{sh}>1$.

The energy density of this radiation field in the host galaxy frame at the location of the spine is
 $U=(1/c)\int_{\Omega_{sh}} I d\Omega=(1/c)I \Omega_{sh}$ ($c$ is the speed of light), as in our approximate treatment the intensity does not have an angular dependence within $\Omega_{sh}$.  Using $\Omega_{sh}=\pi/\Gamma_{sh}^{2}$ we then obtain:
\begin{equation}
U=\frac{\pi I }{c\Gamma_{sh}^{2}}=\frac{ L_{sh, obs}C}{16\pi a R_{sh}^2 c\Gamma_{sh}^{2}}.
\end{equation} 
Recalling that  $\eta=L_{sh, obs}/L_*$ and using equation 
(\ref{eq:lstar})    
 allows us to put this equation in the following form:
\begin{equation}
U=\frac{4 \sigma T_{sub}^{4}}{c}\frac{\eta C }{4 a \Gamma_{sh}^{2}}.
\end{equation}
Note that the first fraction in the above equation is a blackbody energy density and the second fraction is a dimensionless dilution factor.
The energy density in the rest frame of the spine flow plasma can be shown to be $U^{''} =4 U\Gamma_{sp}^{2}$
\citep[following a calculation similar to][]{dermer94}, where double primed variables refer to the co-moving frame of the spine flow:  
\begin{equation}
U''=       \frac{ L_{sh, obs}C}{4\pi a R_{sh}^2 c} 
\frac {\Gamma_{sp}^2}{\Gamma_{sh}^{2}}=
\frac{4 \sigma T_{sub}^{4}}{c} \; \frac{\eta C }{a}    \frac{\Gamma_{sp}^{2}}{\Gamma_{sh}^{2}}
\label{eq:u}
\end{equation}

\subsection{Photon energy density in the spine coming directly from the sheath}

Assuming isotropic emission in the rest frame of the sheath, the energy density of the sheath's synchrotron radiation in its comoving frame is given by:
\begin{equation}
U'_{sh}=\frac{L_{sh}'}{4\pi R_{sh}^{2}c}= \frac{L_{sh,obs}}{16\pi R_{sh}^{2}c\Gamma_{sh}^{3}}, 
\end{equation}
where  we have used eq. (\ref{eq:lums}).
The energy density of this photon field in the spine frame is $U_{sh}''=(4/3) U_{sh}'\Gamma_{rel}^{2}$ \citep{dermer94},
where  $\Gamma_{rel}$ is the relative Lorentz factor between the spine and the sheath, given by $\Gamma_{rel}=\Gamma_{sh}\Gamma_{sp}(1-\beta_{sh}\beta_{sp})\approx\Gamma_{sp}/(2\Gamma_{sh})$.
Using these we can write:
\begin{equation}
U''_{sh}=\frac{L_{sh, obs}\Gamma_{sp}^{2}}{48\pi R_{sh}^{2}c\Gamma_{sh}^{5}}=\frac{4 \sigma T_{sub}^4}{c} \eta \frac{\Gamma_{sp}^2}{12 \Gamma_{sh}^5},
\end{equation}
 where we have used equation (\ref{eq:lstar}).

 \subsection{Comparison of the energy densities in the spine} 
 
  The ratio of the energy density in the spine rest frame due to the photon field received in the spine from the illuminated MT clouds to the photon field directly illuminating the spine by the sheath is  
\begin{equation}
\frac{U''}{U''_{sh}}=\frac{12 C\Gamma_{sh}^{3}}{a}.
\label{eq:uratio}
\end{equation}

In the case of a low but non-negligible covering factor, as is plausible  in this setting 
{ (see \S \ref{section:discussion}),
 we can set $12 C=1$.
 Also, the sheath is constrained  to be substantially slower than the $\Gamma_{sp} \sim10-20$ spine, and still be relativistic, as otherwise we would detect the counter-sheath with VLBI observations. 
A plausible value for the sheath Lorentz factor is
$\Gamma_{sh}=3$.
For these values of $C$ and $\Gamma_{sh}$
the condition for the cloud-reprocessed radiation energy density in the spine   to be comparable to or dominate over that coming directly from the sheath becomes
 $a=R_{out}/R_{sh}\lesssim 27$.  
 For example, adopting  and setting $R_{sh}=0.2 $ pc \citep{macdonald15}, we see that the cloud processed photon energy density dominates if the sheath-spine system is embedded in the MT by less than $\sim 10\,  R_{sh} \sim 2$ pc.  
 In the framework where some  GeV flares come from distances of a few pc and  the clumpy MT extending for a few pc  this condition is plausible.

\subsection{Viability of SSC for powerful blazars} 

We now address whether the GeV emission  of powerful blazars  can be synchrotron-self Compton   \cite[SSC, e.g.][]{maraschi92} emission. To evaluate this, we approximate the emission region with a sphere of radius $R$, permeated by a magnetic field $B$, and moving relativistically with  Doppler factor $\delta$ relative to the observer. Electrons of Lorentz factor $\gamma$ injected in the source at the rate of $Q$ electrons per second produce the observed flux at the peak of the synchrotron and SSC components. With five model parameters and five observables, namely     the peak frequency of the synchrotron component $\nu_s$, the peak frequency of the SSC component  $\nu_{SSC} $,  the peak luminosity of the synchrotron  component $ L_s$, the  Compton dominance $k$ (the ratio of the GeV to synchrotron luminosity), and the variability timescale $t_{var}$ of the gamma ray emission, the system of equations is closed and   the Doppler factor of the emission region is given by the following expression that contains only observables:
\begin{equation}
\delta= 100 \left[\frac{2}{c^{3}B_{cr}^{2}}\frac{L_{s,46}\,\nu_{ssc,22}^{2}}{\nu_{s,13}^{4}\,t_{var, 1 d}^{2}\,k_2}\right]^{1/4} \label{eq:ssc}
\end{equation}
where $B_{cr}=4.4 \times 10^{13} $ G is the critical magnetic field,   $\nu_{s,13}=\nu_s/10^{13}$Hz, $\nu_{SSC,22}=\nu_{SSC}/10^{22}$Hz, $L_{s,46}=L_{s}/10^{46}$erg s$^{-1}$, $k_2=k/100$  and $t_{var,1d}$ is the variability timescale of the gamma ray emission region in units of one day,  all typical values for powerful blazars \cite[e.g.][]{abdo10,bonnoli11}. 
For
the powerful blazars,  on which we focus here,  
the Doppler factor  given by equation (\ref{eq:ssc})
is significantly higher than the typical values  found from superluminal proper-motions studies  \cite[e.g.][]{lister09,jorstad01}.
 In addition, such high $\delta$ values require either a jet with an opening angle of $\sim 1/\delta$ that is extremely well aligned to the observer and therefore with unrealistic de-projected lengths, or a jet with opening angle much greater than $1/\delta$ that would have to be extremely powerful. For these reasons, we disfavor a SSC interpretation of the  GeV emission of powerful blazars. Similar conclusions about the inadequacy of the SSC process for powerful blazars have been reached before \cite[e.g.,][]{abdo10}.

The SSC process can still be important for powerful blazars, as the synchrotron photon energy density  in the spine comoving  frame 
$ U_s=L_{s,46} /(4 \pi c^3 t^2_{var} \delta^6) $
can be a non-negligible fraction of the photon energy $U''$ in the spine due to the sheath radiation reprocessed by the clouds.
For reasonable jet parameters, the peak of the SSC SED is in the hard X-ray regime. For example, using equation (\ref{eq:ssc}) with $\delta=10$ and requiring that the SSC component has comparable power to the synchrotron one ($k=1$), we find that the peak of the SSC component is at $\nu_{SSC}=10^{19}$ Hz, an energy of $\sim 40$ KeV. This is in agreement with modeling  of the SEDs of powerful blazars \cite[e.g.,][]{boettcher13}.

 \section{An example SED }\label{section:SED}
 
  We now apply the above scenario to evaluate if the resulting  SED from the spine compares well to that of high-power blazars.
  The SED of powerful blazars is characterised by two spectral components, the first peaking at sub-mm to IR and the second below/around  $\sim 100$ MeV, with the high energy component dominating in apparent luminosity by $\sim 10-100 $. In the context of leptonic models there is significant contribution or even dominance of SSC at the X-ray band \citep[e.g.,][]{sikora94}.   

To simulate  our proposed scenario we adopt
 a covering factor $C=0.1$, a sheath plasma Lorentz factor $\Gamma_{sh}=3$ and a ratio $a=R_{out}/R_{sh}=3$. 
 With this set of parameters, using equation (\ref{eq:uratio}) we find that the energy density in the spine's rest frame due to the MT is a factor of $10$ higher than the energy density of the radiation coming directly from the sheath.
 We then use the above parameters in  equation (\ref{eq:u})  to find $U''= 1.2 \times 10^{-2}$ erg cm$^{-3}$ by setting
$T_{sub}=1200$ K,  $\eta=L_{sh, obs}/L_*=1/2$, and $\Gamma_{sp}=20$
(note  that at distances of several pc the energy density of the accretion disk and BLR are at least $\sim \Gamma_{sp}^4$ lower as their photons are entering the spine from behind).
 To obtain a Compton dominance of $\sim  30$, we require $B= (U''/(30\times  8 \pi))^{1/2}=0.1$ G.

 \begin{figure}[t]
 	\epsscale{1.1}
	\plotone{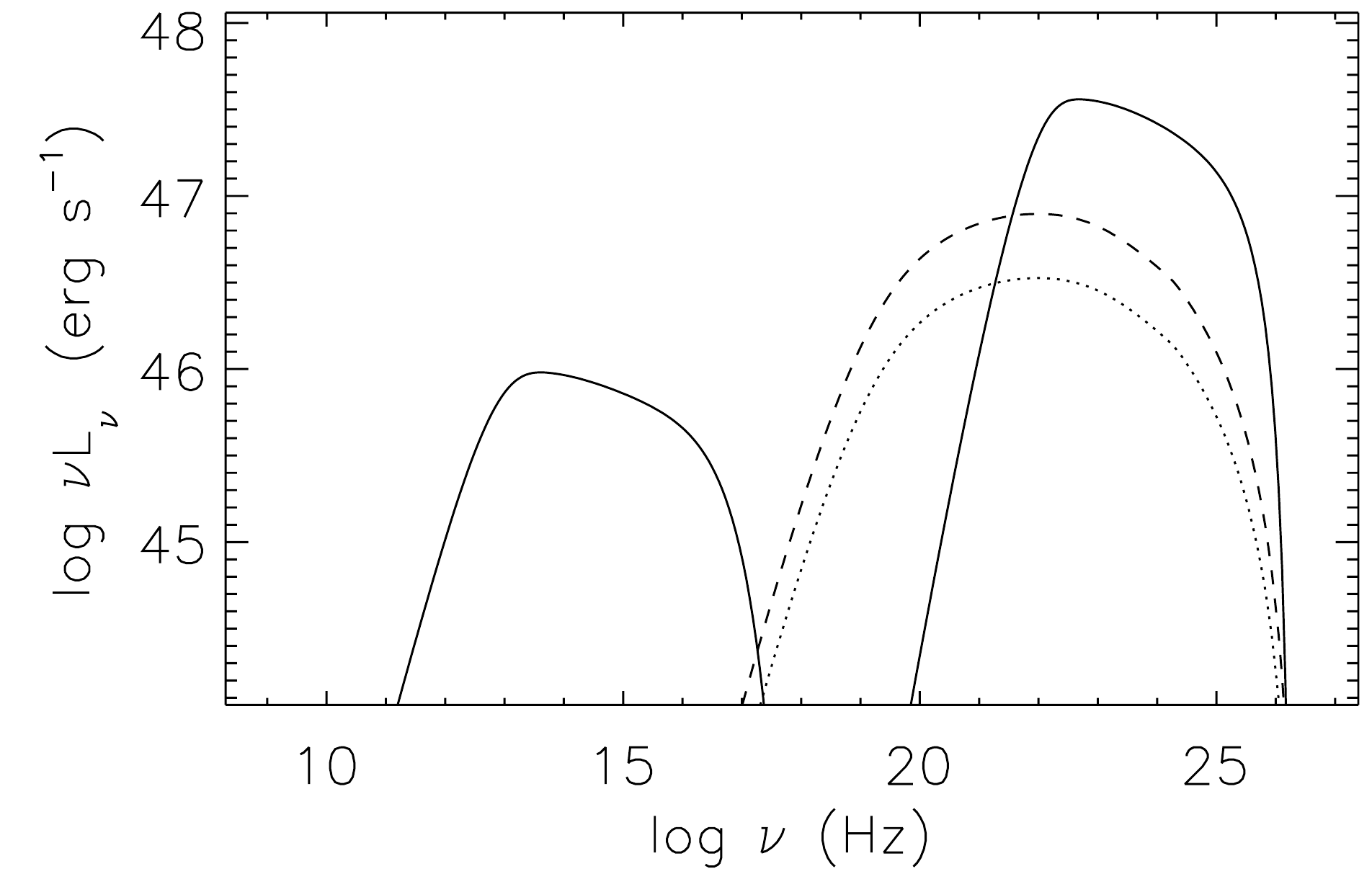}
	\caption{The SED resulting from the parameter values motivated in \S \ref{section:SED}. The low frequency solid line is the spine synchrotron SED, while the high frequency solid line is the spine IC emission from seed photons coming from the MT that is heated by the sheath emission. The broken line is due to SSC within the spine. We also anticipate spine IC emission with the seed photons being sheath synchrotron photons directly entering the spine. As discussed,  for this component we anticipate a level of $\sim 10$ below the IC component resulting  from seed photons coming from the MT, but its exact spectral shape depends on the unobservable synchrotron emission of the sheath. For demonstration purposes, we plot this component as a dotted line, assuming an SED similar to the spine's SSC component but with a peak luminosity $\sim 10$ times lower than the spine IC emission resulting from the MT seed photons.}
	\label{fig}
 \end{figure}

The spectrum of  the reprocessed by the MT radiation of the  sheath will be a superposition of black bodies with the hottest coming from the innermost radius $R_{sh}$ and the coolest coming from the outermost radius of the MT $R_{out}$, with $T(r)=(L_{sh, obs}/16 \pi \sigma r^2)^{1/4}$. Using this and equation (\ref{eq:lstar})  we obtain
$T(R_{sh})=\eta^{1/4}T_{sub}$ and $T(R_{out})=a^{-1/2}\eta^{1/4}T_{sub}$. This means that for small values of $a$, as in this example where $a=3$, the temperature ratio $T(R_{sh})/T(R_{out})=a^{1/2}$, is not a large number and we can approximate the SED of the sheath radiation with a single black body function at 
$T_{eff}=[T(R_{sh}) T(R_{out})]^{1/2}$  and energy density in the spine rest frame given by equation (\ref{eq:u}),
which can now be written as
\begin{equation}
U''=\frac{4 \sigma T_{eff}^{4}}{c} 
     \frac{C\, \Gamma_{sp}^{2}}{\Gamma_{sh}^{2}},
\end{equation}
an expression that absorbs $\eta$ and $a$     in the definition of $T_{eff}$.

To reproduce typical variability timescales of a few  hours to a day, we set the spine radius $R_{sp}=2 \times 10^{16}$ cm. The electron energy distribution (EED) injected in the spine is a power law of index $p=2.5$ confined between electron Lorentz factors $\gamma_{min}$ and $\gamma_{max}$.  Because the system is in the fast cooling regime, the peaks of the synchrotron and IC emission are produced by electrons of Lorentz factor $\sim \gamma_{min}$. Requiring a synchrotron peak at $\nu_s\approx 10^{13}$ Hz, sets $\gamma_{min}=10^3$. The requirement that the synchrotron mechanism cuts off before the X-rays is satisfied with  $\gamma_{max}=10^5$. The comoving injected power is set by  requiring a  blazar  GeV luminosity of $L_{GeV}\approx 5 \times 10^{47}$ erg s$^{-1}$,  as seen in bright GeV blazars \cite[e.g.][]{abdo10}. Using an  equation similar to equation (\ref{eq:lums}), we find $L_{inj}=L_{GeV}/(4 \Gamma_{sp}^3)=2 \times 10^{43}$ erg s$^{-1}$. Using the  parameter values we just motivated, we plot in figure (\ref{fig}) the SED produced by the spine.    Our single-zone code   applies an implicit  numerical scheme for solving the electron kinetic equation similar to that of \cite{graff08}, first introduced by \cite{chang70}.
Our code  follows the radiative losses in the injected EED and uses  the full  Klein-Nishina cross section for IC scattering energy losses and emission calculations.

\section{Conclusions and Discussion} \label{section:discussion}

Recent mutiwavelength campaigns  \citep[e.g.][]{marscher10} strongly suggest that a fraction of the {\sl Fermi} observed blazar flares take place a few to several pc from the central engine. At these distances there is no significant external photon field  for producing the GeV emission 
of observed blazar gamma-ray flares
via IC scattering from jet relativistic electrons: the BLR is confined within the IR-bright inner part of the MT,
which in turn does not exceed a distance of $\sim 1$ pc from the central engine \cite[e.g.][]{koshida14,nenkova08}  for the powerful sources under consideration. A plausible solution is 
a spine-sheath geometry for the emitting region (\cite{macdonald15}, but see \cite{nalewajko14} that argue that this mechanism would become relevant only for observed sheath luminosity that would rival that of the spine).

Here we suggest another seed photon mechanism.
We start by adopting a picture of the MT that extends for a few pc beyond the IR-emitting inner radius  and has a dust cloud angular distribution that extends with a diminishing density to small polar angles \cite[e.g.,][]{nenkova08}. 

We then show that for  a spine-sheath jet configuration located within the MT there is   a reasonable part of parameter space  in which the seed photon energy density in the spine is dominated by sheath photons that have been reprocessed  by the dust clouds within the wide opening angle of the sheath synchrotron radiation.
This differs from the model of \cite{ghisellini96}, as it avoids the problem of requiring reprocessing clouds to be within the very small opening  angle of the spine radiation.

We explore now  plausible values of the AGN covering factor at low polar angles.
Recent high resolution mid-IR observations and modeling of nearby quasars  \citep{martinez17} with the clumpy  molecular torus model \citep{nenkova08}, show that non-negligible covering factors are plausible at small polar angles: for a cloud distribution $N=N_0\, Exp[{-(\theta-90^o)^2/\sigma^2}]$, where $N_0$ is the number of clouds encountered by a line of sight in the equatorial direction, $\sigma$ is the $1/e$ angullar opening of the clumpy torus and $\theta$ is the angle of the line of sight to the polar direction, the covering factor of the clouds at $\theta$ is $C=1-e^{-N}$. Adopting $N_0=5$,  $\sigma=30^\circ$, 
$\theta=30^\circ$,  within a wide permitted range \citep[see table 11 of][]{martinez17}, we obtain a covering factor of $0.1$. This shows that a clumpy molecular torus can provide a non-negligible covering factor at low polar angles.

Variability in our model can result from a range of disturbances in the system. We consider here two types of variations in the injected EED.
In the first case the injected EED amplitude increase
takes place only in the sheath. Then, for both the spine-sheath-only model and the spine-sheath embedded in the MT model  the synchrotron emission of the spine is not expected to vary significantly,
as neither the spine magnetic field, nor the spine EED varies. This would result in GeV orphan flares as in blazar PKS 1222+216 \citep{ackermann14}.  
A difference between the two models that could be used to discriminate between them, is that while in the spine-sheath embedded in the MT model only the GeV part of the high energy component should vary, in the spine-sheath only model the hard X-ray to MeV flux would also vary  with similar amplitude.
This is because the MT-embedded spine-sheath model (spine-sheath only model) seed photons have a narrow (broad) spectral distribution, and this is reflected in the energy width of their IC spectra. 
Consider now the case where in the  MT-embedded spine-sheath model the maximum energy of the EED increases in both the spine and the sheath. 
In this case the luminosity of both the spine and the sheath increases and it is possible that
dust is sublimated within the sheath radiation opening
angle (that would be the case when $L_{sh,obs}>L_\star)$. If the maximum EED energy increases sufficiently, the synchrotron production of UV ionizing photons will illuminate the  clouds  from which the dust has been sublimated, which in turn
will produce line emission  \cite[we think of such clouds as parts of the polar part of the MT, possibly parts of an outflow, e.g., ][]{netzer15}. This line emission would temporally correlate with the optical-UV variations
of the spine and sheath synchrotron emission. Such correlations between the optical-UV continuum and emission line variability has been  tentatively detected by  \cite{isler13} and \cite{leon-tavares13}  in the blazar 3C 454.3 and very recently reported with high statistical significance
by \cite{jorstad17} in the blazar CTA 102.

 \acknowledgements{MG acknowledges support from Fermi grant NNX14AQ71G}

\end{document}